# Magnetohydrodynamic Operating Regimes of Pulsed Plasma Accelerators for Efficient Propellant Utilization

Ethan Horstman*, Adrian Woodley*, and Thomas C. Underwood

Department of Aerospace Engineering and Engineering Mechanics, The University of Texas at Austin, Austin, USA
E-mail: thomas.underwood@utexas.edu
*These authors contributed equally to this work

## Abstract

The presence of magnetohydrodynamic (MHD) acceleration modes in pulsed plasma thrusters has been verified using the magnetic extension of Rankine-Hugoniot theory. However, the impact of initial conditions within the accelerator volume on the formation and structure of these modes remains poorly understood. This work develops a regime map to clarify how key initial conditions—such as propellant gas dynamics, pulse energy, and the timing between propellant injection and discharge initiation—govern transitions between two distinct MHD operating modes, a magneto-detonation and magneto-deflagration, along with an unstable transition regime that connects them. To characterize these regimes, a combination of time-of-flight and thrust stand diagnostics was used to assess their properties, scalability, and structure while operating with air. Time-of-flight measurements reveal that reducing the initial downstream propellant mass ($m_{dwn}$) of air from 120 µg to 60 µg shifts the thruster from the magneto-detonation to the magneto-deflagration regime, increasing exhaust velocity ($v_{ex}$) from 20 km/s to 55 km/s. In this regime, the thruster exhibits improved propellant utilization as less mass is injected. At a constant 8 kA of peak current, specific impulse ($I_{sp}$) increases from ~100-2000 s as $m_{dwn}$ decreases from 70 to 10 µg, corresponding to an increase in utilization efficiency ($\eta_{\text{util}}$) from 5% to 35%. Thrust-to-power ratios, measured using a thrust stand, also improve with peak current in the magneto-deflagration regime, increasing from 4.5 mN/kW to 8 mN/kW and 6.7 mN/kW for injected mass bits of 25 µg and 50 µg, respectively. Operating in the magneto-deflagration mode enhances $I_{sp}$ and $v_{ex}$ significantly while improving $\eta_{\text{util}}$. This work provides critical insights into how the initial conditions in pulsed plasma thrusters dictate the formation of ionization waves, structure of plumes, and the performance of thrusters. These insights show how pulse shaping methods can be used to optimize the performance of pulsed plasma thrusters in environments where high $I_{sp}$ and $\eta_{\text{util}}$ are crucial.

Keywords: Magnetohydrodynamics, Plasmas, Electric Propulsion, Air-Breathing Electric Propulsion, Pulsed Plasma Thrusters, Plasma Waves, Molecular Propellants

## 1. Introduction

The use of transient electromagnetic fields is a mechanism to accelerate plasma flows to high exhaust velocities. Studies have shown that coaxial inductive accelerators can propel monoatomic propellants, such as Argon (Ar), to super-Alfvénic speeds (i.e., ~100 km/s) using the Lorentz force (1–4). Similar performance has been demonstrated in pulsed accelerators that operate with molecular propellants, including diatomic species (e.g., $H_2$, $N_2$, etc.) and reactive mixtures (e.g., air) (5,6). However, challenges with propellant utilization and electrode degradation have limited the scalability of electromagnetic acceleration, and particularly transient inductive thrusters within electric propulsion (EP) (7). Recent advances in self-refuelling architectures, such as air-breathing electric propulsion (ABEP), create a new design opportunity that favors systems which are capable of achieving high exhaust velocities—a regime where pulsed inductive plasma thrusters offer operational advantages (8,9). Yet, questions remain regarding the dynamics and acceleration mechanisms in these devices, namely how discharge pulse shaping, energy transfer mechanisms, and propellant gas dynamics influence the structure and scalability of electromagnetic acceleration. Understanding these factors is crucial for the development of pulsed accelerators that operate efficiently on complex molecular propellants.

Existing electromagnetic devices have been shown to feature high exhaust velocities while operating with small quantities of molecular propellant (down to ~10 µg) (10,11). This includes thrusters that operate in pulsed modes (i.e., inductive or pulsed plasma thrusters (PPTs)) or continuous modes, using either induced or applied magnetic fields (10,12–26). Electromagnetic devices have been shown to achieve $I_{sp}$ higher than electrostatic (up to 10,000 s and 5,000 s, respectively) but with lower overall thrust efficiencies (~ 10% and ~ 50%, respectively) (Fig. 1). Despite the lower thrust efficiency, electromagnetic thrusters are advantageous in applications that require low propellant flow rates, as they can function with smaller mass bits. The increasing interest in using alternative molecular propellants, particularly in low flowrate conditions, highlights operational scenarios where traditional thrusters, such as Hall-effect or quasi-steady devices more broadly, face significant challenges. Nonetheless, scalability constraints in electromagnetic devices and PPTs more specifically—including limitations in maximum power and current, inefficient propellant utilization, thermal losses, and electrode degradation—continue to restrict their feasibility for practical mission designs (27–30).

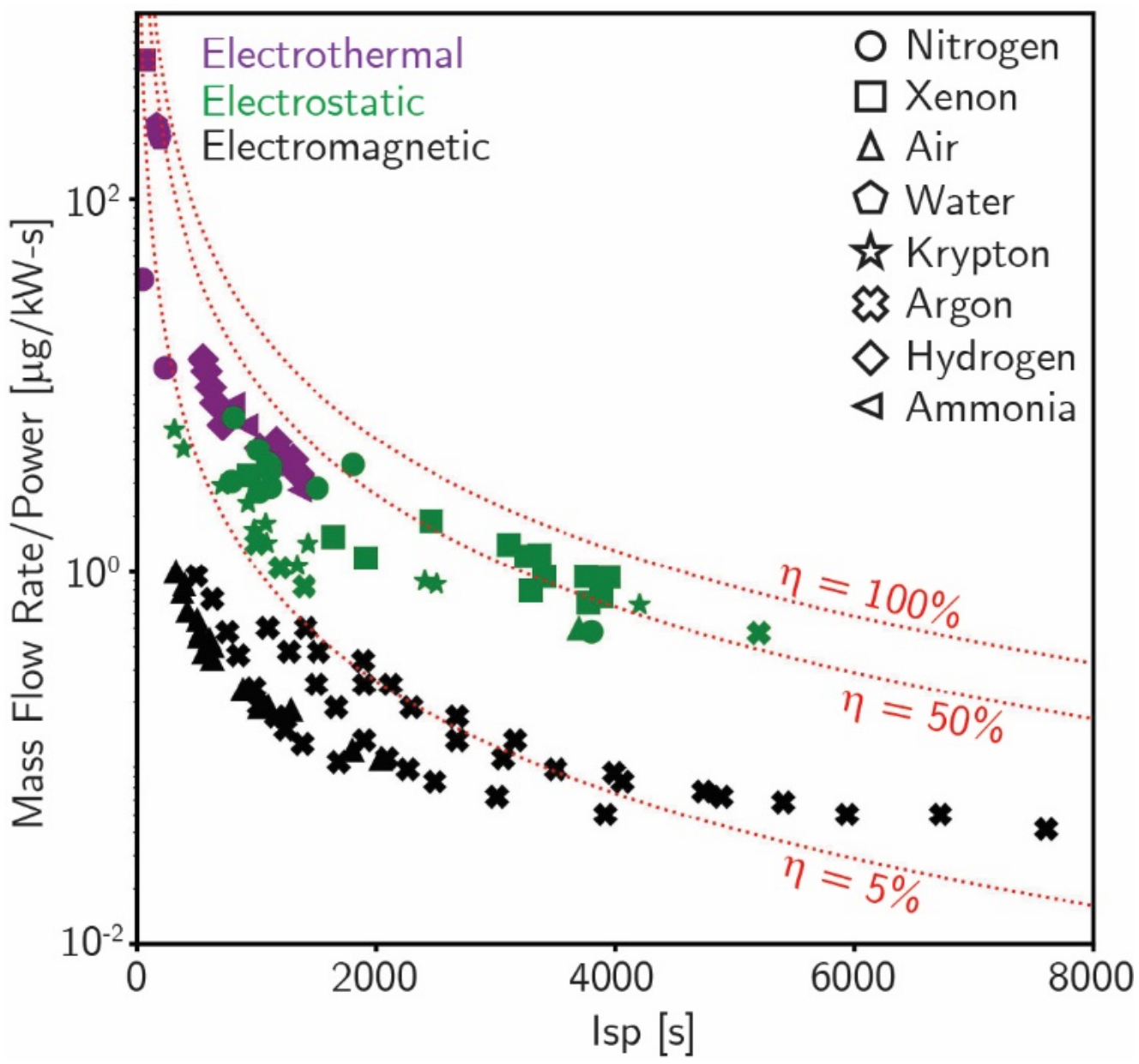

**Figure 1.** The performance of electromagnetic, electrothermal, and electrostatic acceleration mechanisms clusters based on the specific impulse ($I_{sp}$), mass flow rate, and input electrical power for a variety of propellants. Electrothermal and electrostatic architectures feature higher thrust efficiency ($\eta = TI_{sp}g/2P$) across atomic and molecular propellants. Electromagnetic architectures feature higher $I_{sp}$ and operability with molecular propellants including air. Data taken from (10,12-26).

Recent advances in understanding transient states in electromagnetic thrusters have revealed distinct acceleration modes and new strategies for improving mass utilization and overall thrust efficiency. These mechanisms manifest as two principal ionization wave modes: (1) a transient, propagating mode, characterized by a localized plasma current sheet that advects through the discharge volume, entraining and ionizing neutral species, and (2) a quasi-steady mode, wherein discharge current is spatially distributed throughout the acceleration volume (31). The existence of these modes is predicted by the magnetized extension of the Rankine-Hugoniot jump conditions (1,11). While theoretical models provide a foundational understanding for the properties of these wave structures once they form, significant gaps persist in characterizing their evolution, regime boundaries, mode transitions, and their performance implications for thruster efficiency. Further refinement of predictive frameworks, coupled with experimental validation, is essential for leveraging these wave phenomena in thruster optimization strategies. In particular, advanced control techniques, such as pulse shaping and precise propellant injection, could enable control or even modulation of the acceleration mode, thereby enhancing performance and operational flexibility.

In this work, the mechanisms governing the formation of distinct magnetohydrodynamic acceleration modes, and their implication on performance in PPTs, are investigated experimentally. We demonstrate the existence of these modes and systematically characterize their transitions by varying key initial conditions, including propellant gas dynamics, pulse energy, and the timing between propellant injection and discharge initiation, within a defined acceleration volume. Our measurements reveal a regime boundary between two distinct acceleration structures, along with an unstable transition region, both of which depend on operating conditions and propellant loading when an ionization wave forms. Furthermore, the performance characteristics of each mode is analyzed through measurements of mass utilization, exhaust velocity, and thrust-to-power ratio. Our results show that operating PPTs at low propellant mass flow rates, combined with high discharge currents, produces plumes with higher exhaust velocity and improved mass utilization while reducing macroscopic instabilities. This work represents an initial step toward integrating control strategies within PPTs by examining how initial conditions and energy deposition influence acceleration dynamics.

## 2. Pulsed Inductive Thrusters

### *2.1 Requirements of Electromagnetic Acceleration Architectures*

Air-Breathing Electric Propulsion (ABEP) has emerged as a promising technology for station-keeping in very low Earth orbit (VLEO). This approach mitigates the high drag forces at low altitudes (i.e., < 450 km) by utilizing atmospheric gases as propellant, either fully (complete air-breathing) or partially, reducing the need for onboard propellant storage (10,32,33). To be effective, ABEP thrusters must achieve thrust-to-power (T/P) ratios high enough to offset drag using the available atmospheric propellant. This imposes performance requirements that depend on spacecraft geometry, altitude, and the efficiency of the capture and collection processes within the ABEP system. In the free-molecular regime, the amount of propellant that can be harvested and the minimum $I_{sp}$ required for sustained operation ($I_{sp,req}$) are determined by environmental conditions and system design,

$$\dot{m} = \eta_{col}\sigma\frac{F_D}{P_{sol}}A_{in}, \qquad (1)$$

$$I_{sp,req} = \frac{F_D}{\dot{m}\eta_{col}\eta_{util}g_0}, \qquad (2)$$

where $\dot{m}$ is harvested mass flowrate, $A_{in}$ is the inlet area, $F_D=0.5\rho C_D A_{drag} v_{orbit}^2$ is the drag created by the spacecraft geometry in the limit of free-molecular flow, $C_D$ is the drag coefficient (assumed to be 2.2), $\eta_{col}$ and $\eta_{util}$ are the propellant collection and utilization efficiencies, respectively, and $\sigma=\eta_{sol}Q_{sol}A_{sol}/C_D v_{orbit} A_{drag}$ is an effective drag pressure found from spacecraft solar panel and drag geometry. Propulsion systems are assumed to be powered using solar panels that provide $P_{sol}=\eta_{sol}Q_{sol}A_{sol}$, where $A_{sol}$ is the solar panel area ($4A_{in}$), $\eta_{sol}$ is the solar panel efficiency (20%), and $Q_{sol}$ is the solar flux (1367 W/m$^2$ in Earth orbit). These relationships constrain thruster design and dictate the feasible acceleration mechanisms for low-altitude operation, where satellites travel at orbital velocities of ~7.5 km/s ($v_{orbit} = \sqrt{\mu/R}$ for circular orbits, where R is the orbital radius and μ is the gravitational parameter for Earth). For instance, as altitudes reduce from 400 to 150 km, increases in atmospheric drag cause the $F_D/P_{sol}$ experienced by satellites to increase from ~ 0.10 μN to ~ 35 mN for a drag facing area of 0.3 $m^2$ (Fig. 2). However, the rate that propellant can be harvested at only increases from ~ 2 μg/s to ~ 4 mg/s over the same altitude range for an inlet area of 0.3 $m^2$. This presents a significant challenge for widely deployed thruster architectures, such as electrostatic and electrothermal systems. While these technologies can achieve relatively high T/P ratios (~20 mN/kW for Hall-effect Thrusters (HETs) and ~250 mN/kW for arcjets), their high propellant consumption (~2000–6000 μg/s) makes them impractical for fully air-breathing operation (Fig. 2A, (10,21,26,32–37)). These requirements favor acceleration mechanisms that can overcome inefficiencies in the collection ($\eta_{col}$) and utilization ($\eta_{util}$) of propellant by generating high $I_{sp}$ plumes (Fig. 2B). Under fully air-breathing conditions, system inefficiencies drive the minimum $I_{sp}$ of candidate thrusters beyond the levels typically achieved by HETs at ~1100 s and electrothermal thrusters at ~260 s (Fig. 2C) (32,38).

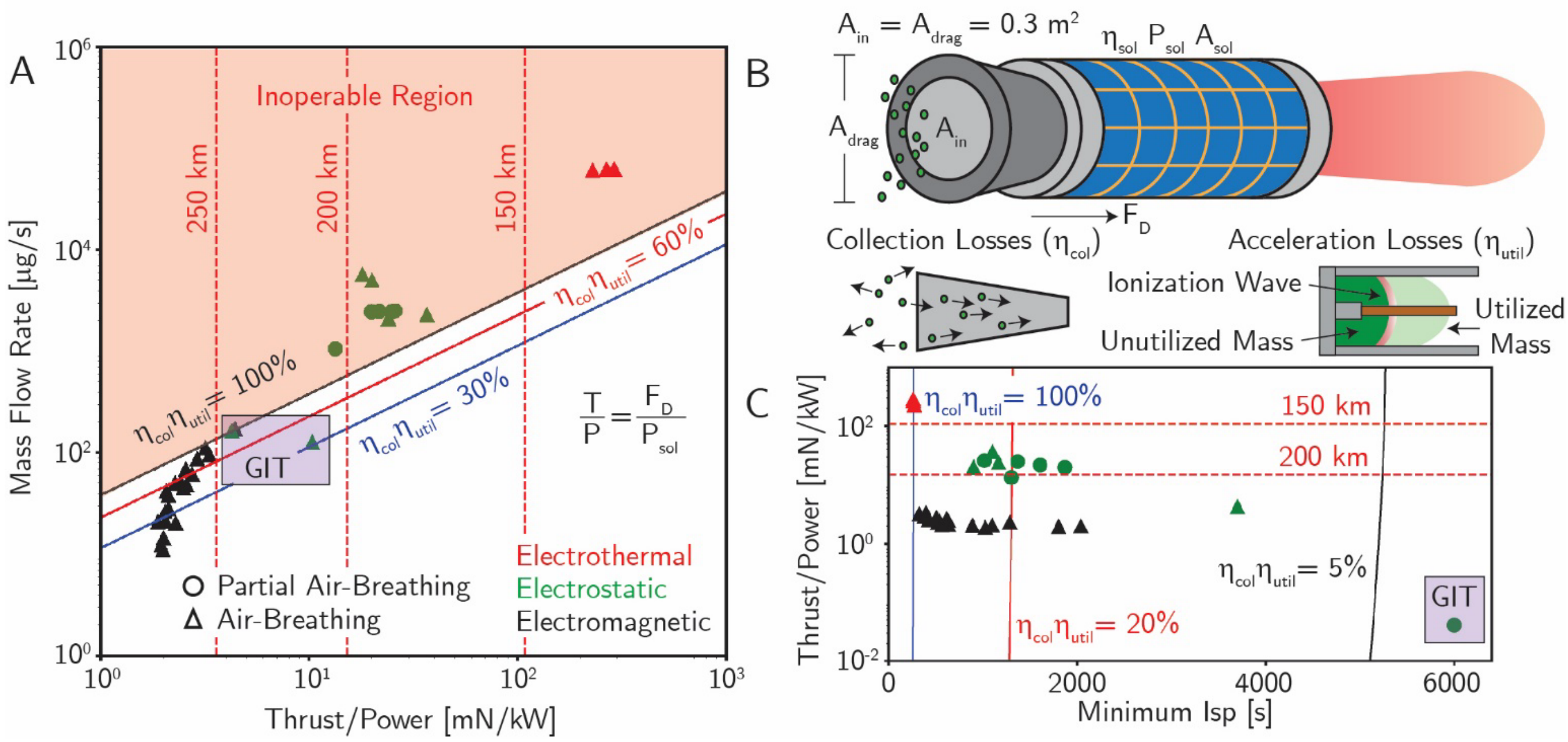

**Figure 2.** Air-breathing electric propulsion devices must achieve high $I_{sp}$ to operate in VLEO where there is a limited amount of propellant that can be harvested from the atmosphere. (A) Thrusters must maintain thrust to power ratios (T/P) for orbital station-keeping using the flow rates of propellant that can be harvested from the atmosphere. Constraints on the collection ($\eta_{col}$) and utilization ($\eta_{util}$) efficiency of propellant impact whether an air-breathing system is inoperable at a given altitude. (C) Inefficiencies in $\eta_{col}$ and $\eta_{util}$ increase the $I_{sp}$ that is required for VLEO station-keeping. Datapoints gathered from (10,21,26,32-37).

Electromagnetic thrusters have demonstrated a capability to generate high $I_{sp}$, operate on complex molecular propellants (e.g., $N_2$, air, etc.), and function in conditions with limited flow rates of propellants. For example, electromagnetic accelerators can achieve exhaust velocities up to 75 km/s while operating on air. PPTs have been shown to generate these velocities while acting on instantaneous mass bits (i.e., of a single firing event) as low as ~10 μg, conditions that are compatible with candidate spacecrafts ($A_{in}$ = 0.3 m) over an altitude range spanning 200-400 km depending on the efficiency of collection, compression, and utilization. Additionally, the periodic thrust generation in PPTs allows for the collection and compression of propellant to be stored and dispensed as needed. This capability helps address the challenge of compressing harvested propellant streams to sustain a plasma (i.e., generate breakdown pressures) while maintaining enough mass flow to counteract drag. For example, passive compression systems have been theorized to be capable of compression ratios of ~1-$10^3$ and collection efficiencies of ~0.3-0.65, compared to $10^5$ and 0.65, respectively, for store-and-dispense active schemes (9). However, performance and scalability challenges remain for PPTs to achieve higher thrust and operating powers while maintaining efficient propellant utilization.

### *2.2 Performance Scalability by Shaping Discharge Conditions*

Controlling discharge conditions and propellant loading in electromagnetic thrusters is one mechanism to enhance their performance. Viability of these thrusters for ABEP demand that they overcome inefficiencies in $\eta_{col}$ and $\eta_{util}$ by sustaining high $I_{sp}$ (> 4000 s), while maintaining harvestable mass flow rates (~10 μg/s) and mass utilization (above 20%), as current densities are increased (Fig. 3A-B). These thrusters exist in various configurations, including gas-fed, ambient-fill, ablative, and magnetoplasmadynamic (MPD) designs, each with distinct propellant injection, discharge features, and performance thresholds (Fig. 3). Ambient-fill thrusters rely on a spark gap to prevent Paschen breakdown while the acceleration volume is filled to a set pressure with propellant. Gas-fed thrusters use precisely timed puff valves to inject gas into the acceleration volume, with discharges occurring once internal pressures reach the Paschen breakdown threshold. Ablative PPTs sublimate solid PTFE pucks to generate propellant. Despite their differences, all electromagnetic thrusters face challenges related to scalability, poor mass utilization, and degradation effects (Fig. 3). MPD and ablative thrusters operate typically at lower peak currents (< 15 kA), whereas gas-fed and ambient-fill designs have been tested at higher currents (40–50 kA) where they have been referred to as plasma guns (Fig. 3A). However, inefficient energy addition in ambient-fill thrusters results in poor mass utilization (< 10%) and low specific impulse (< 4000 s) due to entrainment losses (Fig. 3B). Gas-fed thrusters offer greater control over energy deposition, improving entrainment and utilization efficiencies up to 40%, with recorded $I_{sp}$ values reaching 9000 s using air as propellant (10,39–48). Ablative thrusters achieve up to 60% mass utilization but suffer from degradation effects and limited operational lifespan, making them less viable for VLEO applications.

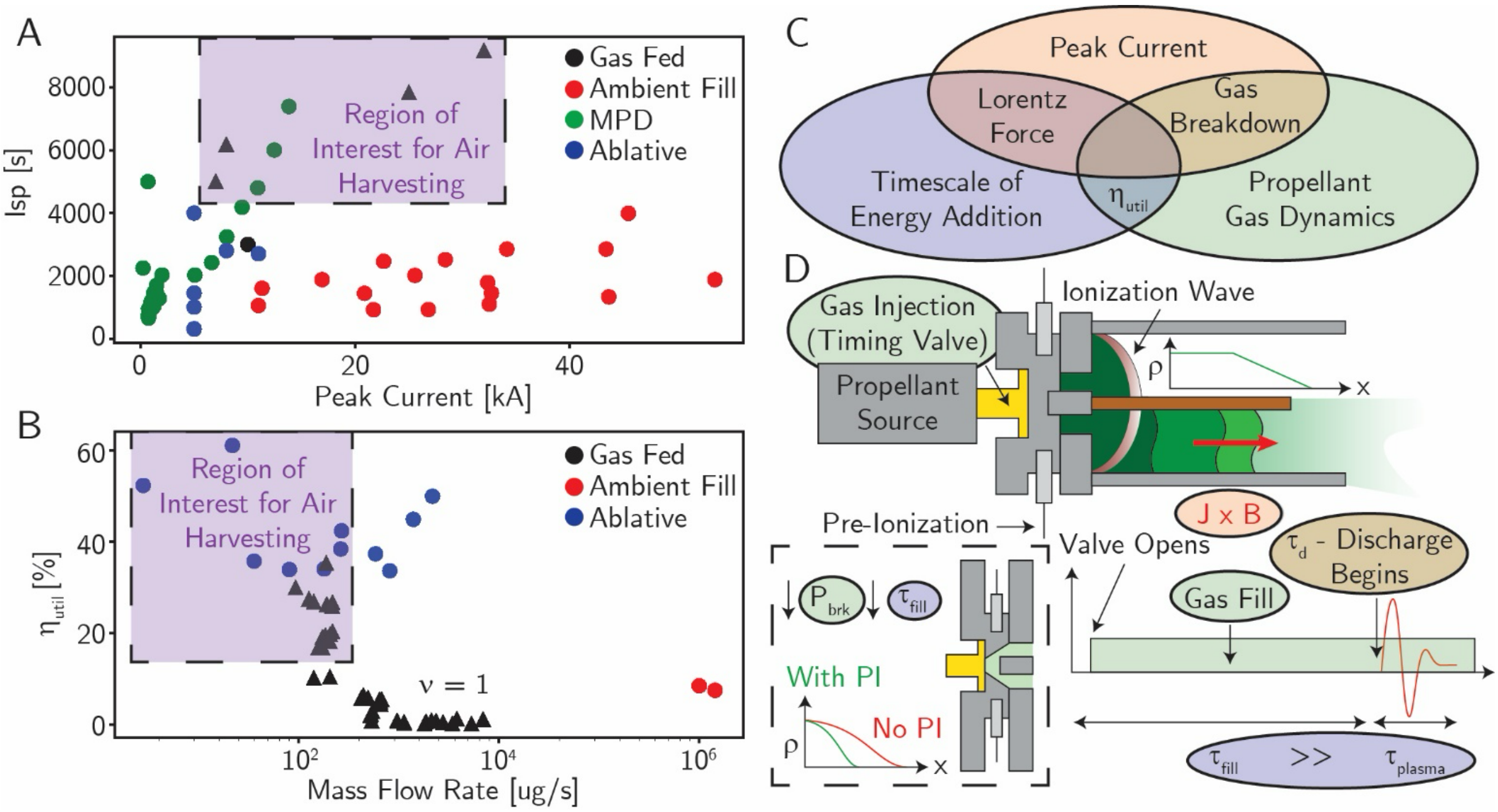

**Figure 3.** (A-B) Electromagnetic thrusters face performance scalability challenges across different architectures, including those using gas-fed, ambient-fill, or ablative propellants, as well as those operating at high or low peak currents (such as pulsed or magnetoplasmadynamic thrusters) and employing either applied or self-field acceleration. Examples include the (A) $I_{sp}$ or (B) mass utilization ($\eta_{util}$) (for gas fed operation). (C) Key operating parameters—peak current, energy addition timescales, and propellant gas dynamics—directly influence $\eta_{util}$ for electromagnetic thrusters. (D) These parameters include the timing between gas injection and discharge initiation (energy addition timescales), the propellant density profile within the accelerator volume (gas dynamics), and the magnitude of the accelerating force on the plasma (peak current). Data taken from (10,39-48).

Thruster performance is also influenced by transient behavior in the electromagnetic acceleration process, which varies based on firing conditions and thruster geometry. This phenomenon is relevant during early transients in MPD devices. However, it is particularly important in PPTs, where the timescales of discharge (~10 µs) are much faster the timescale of gas injection (~1 ms) (Fig. 3D). This mismatch causes difficulties in controlling the energy deposition timescale and pulse shape, which in turn affects propellant entrainment within the accelerated plume. Previous PPT research often neglected these effects, leading to operation in an overfed regime and mass utilization as low as 5%, due to inefficient ionization or excessive propellant outside the acceleration volume as ionization waves form.

Precise control over acceleration dynamics of plasmas in electromagnetic thrusters can be achieved by manipulating the initial conditions within the accelerator volume. This includes variations in how propellant is injected, how discharges are initiated, and the timescales over which energy is added into the propellant (Fig. 3C). Operation of PPTs without consideration of their initial condition, including the gas dynamic distribution of propellant within the acceleration volume before they are fired, leads to changes in their acceleration structure, stability spectrum within plumes, and performance. For example, changes to the peak current in a thruster adjusts the magnitude of the acceleration mechanism (i.e., Lorentz force), degree of collimation within a plume, and mitigates the growth of macroscopic MHD instabilities. Propellant gas dynamics affect the breakdown pressure and gas filling of a thruster, which is a key parameter for the efficient utilization of propellant in PPTs. The timescale and pulse shape of energy addition affects the structure of the acceleration process and the entrainment of mass into the ionization wave. Adding energy before enough propellant has flowed into the acceleration volume can result in the generation of vacuum arcs, while late energy addition can lead to inefficient ionization and acceleration of an overfed volume.

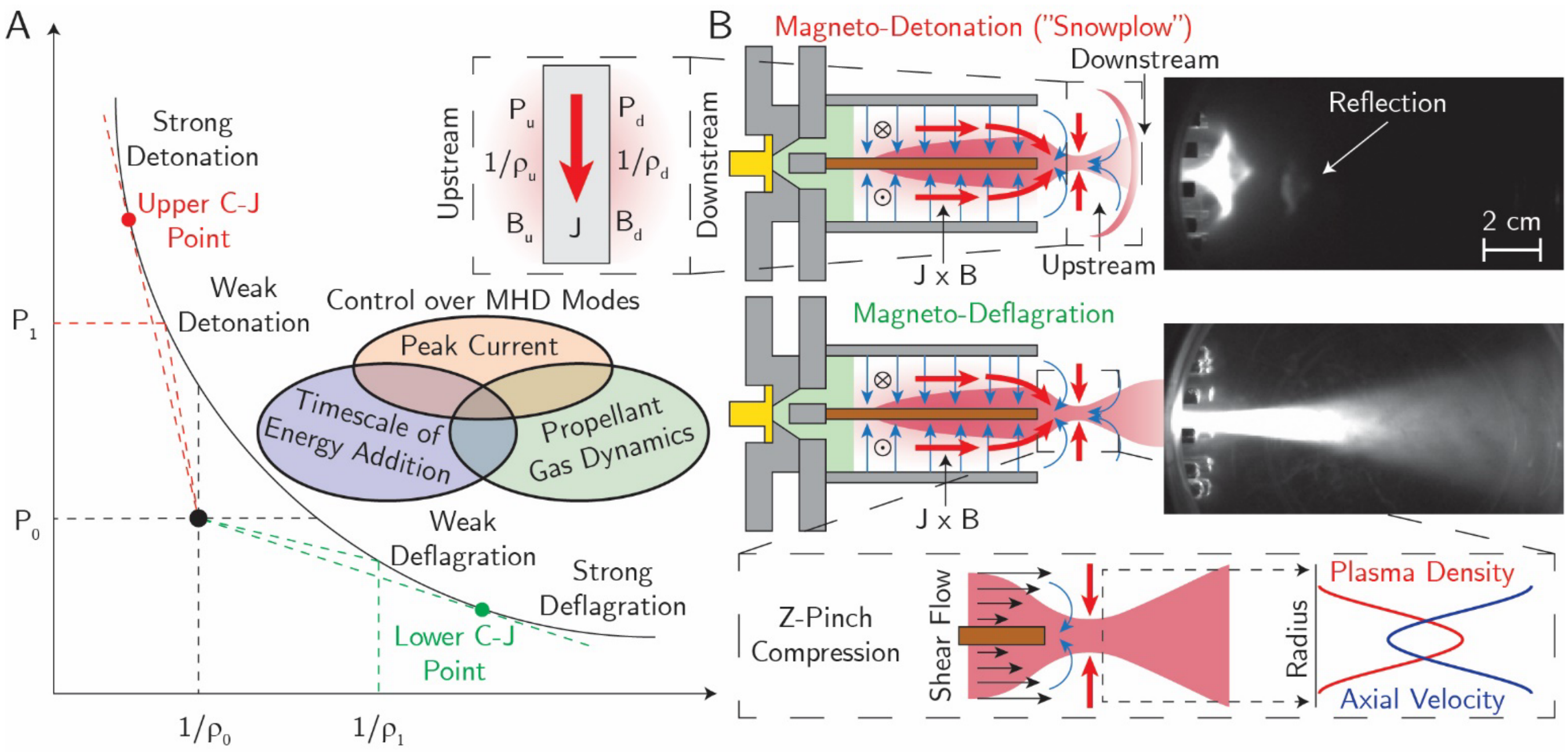

**Figure 4.** (A) The operating mode of a PPT is determined by the thruster's initial conditions, including propellant distribution, the amount of applied energy, and the rate at which energy is delivered to the propellant. Longer pulse duration accelerators are also dictated by these modes during start up transients. Two distinct ionization waves can form, with their properties described using a magnetic analogue of the Rankine-Hugoniot jump relations. (B) These modes, called magneto-deflagration and magneto-detonation waves, feature different structure, current localization, and propellant utilization. The magneto-deflagration mode is distributed throughout the acceleration volume, quasi-steady, and depending on the electrode geometry, can sustain a shear-stabilized z-pinch at the exit plane.

*2.3 Magnetohydrodynamic Operating Modes*

Transient electromagnetic thrusters, such as PPTs, operate in distinct magnetohydrodynamic modes that influence mass utilization, thrust efficiency, and the potential for performance scaling. The existence of these modes has been theorized by observing differences in plume structure under different firing conditions. For example, early measurements of gas-fed PPTs tested them in a pre-filled configuration, where propellant filled the acceleration volume before a discharge was initiated. In these operating conditions, an idealized current “snowplow” formation has been used to describe the structure of the plasma as it propagates down the acceleration volume. Berkery and Choueiri provided evidence of permeability of current sheets within accelerators under these conditions, identifying inefficiencies in propellant entrainment and refining the understanding of the acceleration process (49). Around the same time, Ziemer observed anomalous firing modes in PPTs, where mass utilization and thrust efficiency increased as less propellant was injected (22). Further work by Loebner et al. probed the exit plane of gas-fed PPTs in these operating conditions and found evidence for hyperthermal plasma jets forming when anomalous improvements in efficiency were noted. These jets were found to be stable with increasing compression as more current was supplied, consistent with what is expected in a z-pinch.

The existence of distinct acceleration modes within PPTs has been generalized using a magnetic analogue to the Rankine-Hugoniot (RH) theory (50). The RH theory predicts the existence of two MHD modes that correspond to the upper and lower branch of RH jump conditions (Fig. 4A). These MHD modes have been observed in astrophysical contexts for a century, where plasma jets have been observed to reach megaparsecs in length (51). The upper branch solution (called a magneto-detonation) is noted to have pressure and density increases across the wave. In the magneto-detonation mode, a localized current sheet forms as plasma propagates downstream toward the thruster exit. This "snowplow" effect compresses and sweeps out downstream mass, where the acceleration volume contains neutral gas. As the ionization wave propagates into downstream propellant, it slows, compressing the mass and forming a characteristic shock structure (Fig. 4B). However, the resulting shockwave raises gas temperature, increases electrical conductivity, narrows, and then increases the current density and rate of degradation locally (52). The lower branch wave (magneto-deflagration) produces a quasi-steady ionization wave where pressure and density drop across the wave. In the magneto-deflagration mode, a quasi-steady expansion wave propagates toward the breech end of the accelerator. This mode spreads current across the entire accelerator volume, lowering peak current densities and processing less mass, which can improve propellant utilization. For coaxial geometries, current streamlines also form a magnetic nozzle at the exit plane of the acceleration volume. The nozzle produces a Z-pinch effect, compressing the wave radially as it expands into vacuum (53,54). The presence of these distinct operating modes has been validated previously (55).

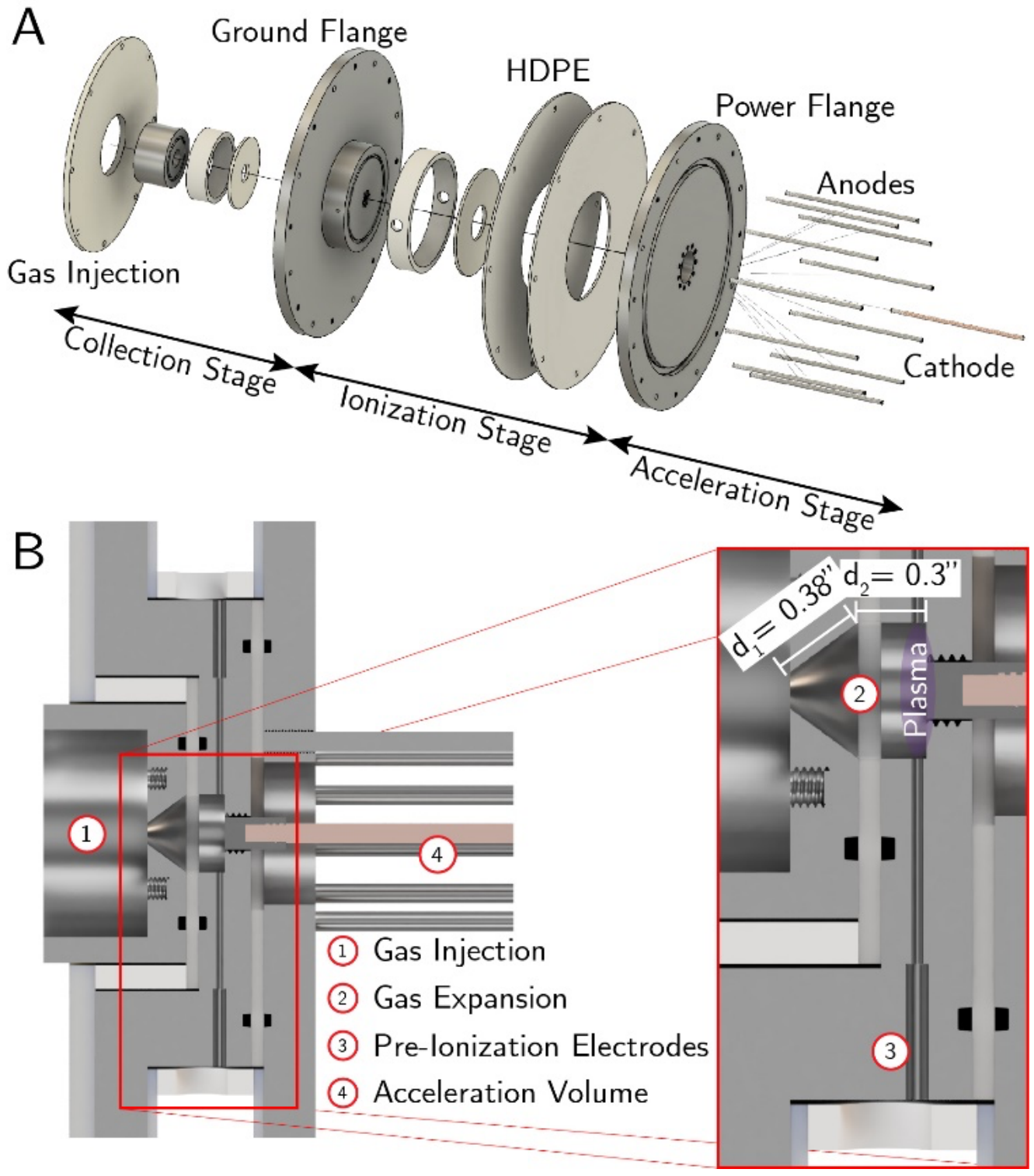

**Figure 5.** (A) Schematic of the gas-fed electromagnetic thruster used in this study. The thruster consisted of a propellant injection, expansion (~0.25 $in^3$ volume with an expansion ratio of 4), ionization, and acceleration stage. The thruster was operated in a variety of firing conditions with air to compare performance and transitions between different MHD modes. The use of an injection valve and pre-ionization scheme enabled control over the density distribution and amount of propellant added to the acceleration stage.

While RH theory predicts the properties of each mode, it does not explain the conditions that lead to their formation, the dynamics of how they accelerate, their performance in thrusters, or how transitions between them can be controlled —key challenges for optimizing PPT performance. This study addresses these gaps by probing the regime boundaries between MHD modes, including how control over the initial condition of each ionization waves dictates mode selection. By altering key parameters in a PPT—such as peak current, energy addition timescales, and propellant injection conditions—the dynamics of the ionization wave can be adjusted to shift between magneto-detonation and deflagration regimes, and the unstable interface that separates them. This framework lays the foundation for improving PPT efficiency, scalability, and applicability in advanced propulsion systems.

## 3. Experimental Setup

### *3.1 Multi-Stage Electromagnetic Thruster*

The UT Austin pulsed inductive thruster is a multi-stage co-axial accelerator that featured separate gas injection, expansion, pre-ionization, and acceleration stages (Fig. 5A). Each stage was integrated into a thruster system to simulate the composition, mass flow rate, and pressure of propellant that can be harvested from altitudes within VLEO. The acceleration stage incorporated a coaxial volume where radial current flow and an induced azimuthal magnetic field accelerated propellant as a quasi-neutral plume, similar to existing self-field electromagnetic architectures. However, by controlling the distribution of propellant within the acceleration volume and precisely timing when energy was added, the thruster was able to transition between different magnetohydrodynamic acceleration modes. Each mode is a different type of current-driven ionization wave that features unique flow structure, propellant utilization efficiency, and thrust scalability for a given input energy.

The thruster prototype was an assembly of different sub-systems that controlled how propellant was introduced, ionized, and accelerated (Fig. 5B). The acceleration stage consisted of two stainless steel (SS) flanges, each 30 cm in diameter, that transited power from an external capacitor bank. High-density polyethylene insulation was added between each flange to eliminate vacuum arcs. Power flowed from the flanges to the acceleration volume that measured 15 cm in length and 5 cm in diameter using a collection of rodded electrodes. This included eight SS anodes and a central copper cathode, all measuring 0.5 cm in diameter. A separate SS baffle (1.5 diameter, 3.8 length) was added to constrain the expansion of propellant through the volume and direct it through the acceleration stage (56).

Propellant was introduced into the thruster using a combination of a fast-acting gas puff valve, expansion volume, and pre-ionization circuit (Fig. 5B). A discharge sequence was initiated by opening a gas puff valve with a 0.8 mm orifice, allowing a desired mass of propellant into the thruster. The quantity of propellant was varied from ~10-400 μg by changing the opening time or plenum pressure of the valve. Propellant then flowed through a 4.1 $cm^3$ volume with an expansion ratio of 4. A pre-ionization circuit was integrated in this volume using a pair of 1.5 mm diameter tungsten rods that, for select MHD operating conditions, were biased at 900 V DC. Propellant then flowed through a series of 10 circularly symmetric orifices (2.5 mm diameter) into the acceleration volume. Seed electrons from the pre-ionization circuit were used to control the breakdown dynamics within the acceleration process and enable control over the initial propellant loading when a breakdown occurred.

The propellant ionization and acceleration processes occurred within the coaxial rodded geometry. The thruster was connected to an optically accessible drift tube (0.6 m long) and a cryogenically pumped high-vacuum chamber to maintain pressures of ~$10^{-7}$ Torr before a discharge sequence was initialized. Characterization efforts within the study focused on single discharge events where the initial conditions (i.e., amount of propellant, distribution of propellant) and discharge parameters (i.e., current, discharge time) could be controlled precisely. Before propellant was introduced into the thruster, the electrodes were connected to a charged capacitor bank along low-inductance transmission lines (~1.5 m long, 70 nH/ft) and floated on the vacuum side of Paschen breakdown. The capacitor bank consisted of four 14 µF capacitors that were connected in parallel. The capacitors were charged between 1 and 4.5 kV in this study, resulting in peak currents of 6-40 kA and pulse durations lasting ~20-40 µs in an underdamped oscillator configuration.

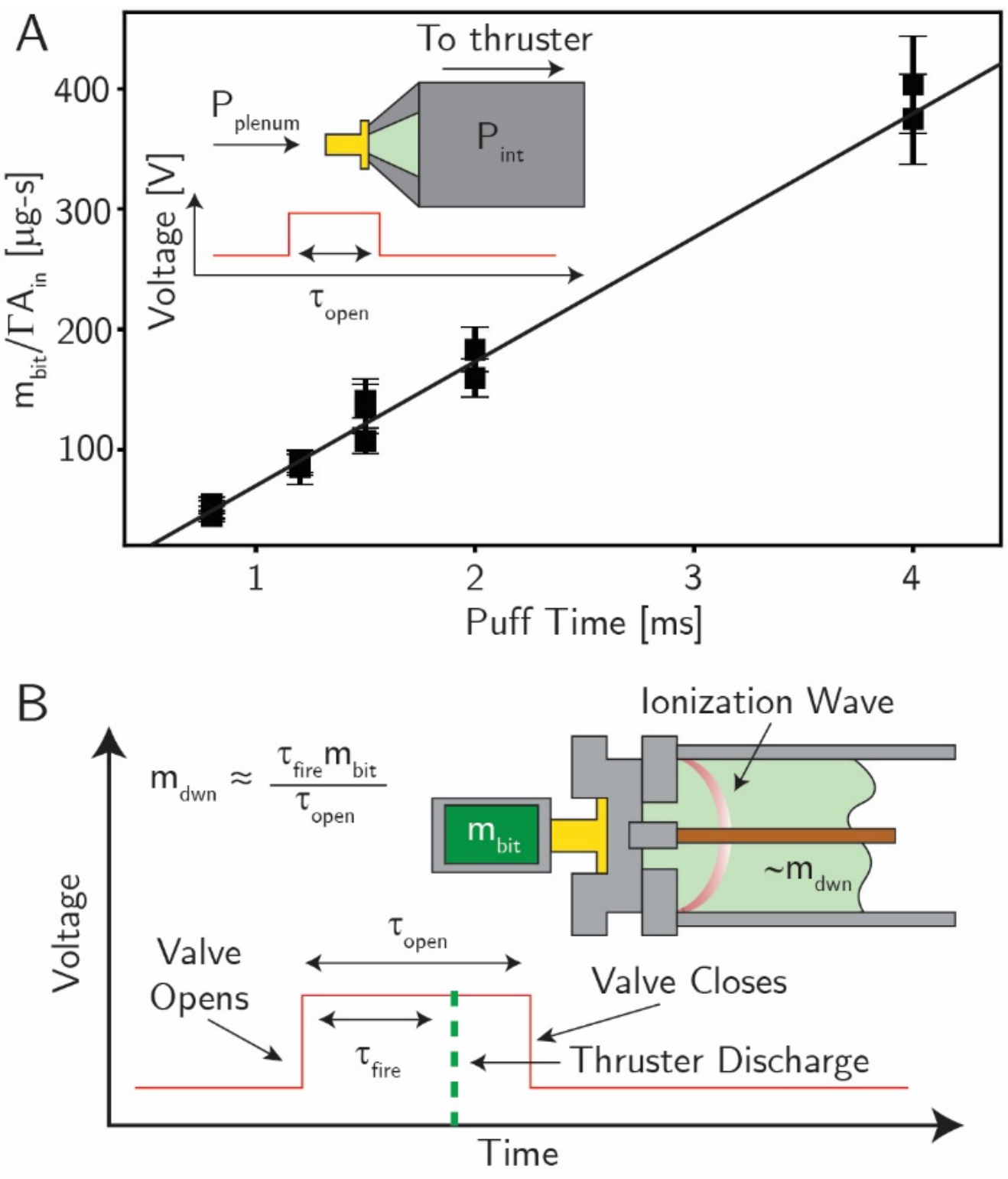

**Figure 6.** (A) Calibration curve showing amount of mass injected per valve opening ($m_{bit}$) for the gas injection mechanism. Mass bits ranging from ~10-200 µg were introduced to the thruster by adjusting valve opening time ($\tau_{open}$) and upstream plenum pressure. (B) The downstream mass ($m_{dwn}$) refers to the amount of propellant present in the accelerator volume at the moment when a discharge is initiated. This metric was quantified by correlating the $m_{bit}$ with the time interval between a valve opening and discharge initiation.

*3.2 Propellant Injection*

Air was injected into the thruster as propellant using a fast-acting solenoid puff valve that leveraged diamagnetic repulsion (57). The valve featured a fixed orifice size of 0.8 mm ($A_{in}$ = 2 mm$^2$) and could be actuated at rates up to 250 Hz. The opening time (500-2000 µs) and plenum pressure (0-75 psig) of the valve were controlled to vary the amount of propellant that was injected into the thruster for a single discharge event ($m_{bit}$ ~10-400 µg). The range of the valve enables replicating propellant harvesting, and amounts of masses that can be collected, at a variety of target altitudes within VLEO. Calibrations between the $m_{bit}$ of air and the properties the valve was operated at (e.g., plenum pressure, opening time) were performed by measuring the pressure rise of a fixed volume after thousands of valve actuations. Results indicate that the $m_{bit}$ scaled linearly as the opening (puff) time or plenum pressure (quantified as a one-way flux, $\Gamma = n\bar{c}/4$, where $n$ is number of gas particles and $\bar{c} = \sqrt{8k_bT/\pi m}$ is the mean thermal speed for particles with mass $m$ and temperature $T$), of the valve increased (Fig. 6A).

The gas dynamics of air determines the propellant utilization for thrusters that rely on Paschen breakdown. When injecting air for a thrust event in the UT prototype, propellant began to fill the thruster after the valve was opened. The amount of time it took for air to fill the volume ($\tau_{open}$) before firing depended on the set voltage between the cathode and anode, whether a pre-ionization circuit was present, and the geometry of the expansion section. Any propellant that remained within the expansion section remained unutilized during the discharge. Once there was sufficient gas in the volume to drive a breakdown, the current discharged from the capacitors over timescales ($\tau_{fire}$), periods that were much shorter than the gas filling time (Fig. 6B). These metrics were measured for each discharge to quantify the propellant distribution and fill fraction of air that was in the acceleration volume as a discharge started (i.e., downstream of an ionization wave). An estimation for this downstream mass ($m_{dwn}$) was generated by weighting the total injected $m_{bit}$ by the ratio $\tau_{open}/\tau_{fire}$ to estimate the amount of air within the acceleration volume during thruster discharge.

*3.3 Impulse Bit: Pendulum Thrust Stand*

A ballistic pendulum thrust stand was used to measure the impulse bit ($I_{bit}$) generated by the thruster across different firing conditions (Fig. 7A). The pendulum featured a circular scoop that was 4.5 cm in diameter, featured a wall thickness of 0.5 cm, and was placed 7 cm downstream of the cathode to collect the accelerated plume as it departed the acceleration volume. The stand was mounted at one end using a flexural pivot, allowing the pendulum to swing freely with a fixed total spring constant ($k_{eq}$) and arm length of 10 cm. Two different flexural pivots with torsional stiffnesses were used: a high-sensitivity pivot ($k_1$ = 0.8 Nm/rad) for low-energy charging conditions (~25-95 J), and a more rigid pivot ($k_2$ = 21 Nm/rad) for higher energies (95-400 J). A Michelson interferometer with a stabilized He-Ne laser (λ = 633 nm, coherence length ~100 m) was paired with the thrust stand to measure its displacement. Interference fringes were generated when the pendulum was displaced by integer multiples of λ/2 during plume impingement. The interference fringes were detected by a photodiode and counted using a peak detection algorithm to infer the pendulum's displacement. A piezoelectric modulation ($f$ = 8 kHz) was used in the reference arm of the interferometer to integrate phase locked detection, mitigate noise, and improve the signal-to-noise ratio of the diagnostic.

The pendulum system was calibrated in vacuum conditions using a voice coil—a well-known technique for such diagnostics, with the procedure detailed elsewhere (34-37). The coil had a diameter of 1 inch, N = 200 turns, and was actuated using a signal generator that could supply voltages up to 10 V for pulse durations between 1 μs and > 100 s. A ring magnet was placed on the back of the pendulum to induce a magnetic repulsion force and calibrate against a known impulse. Magnetic forces from this interaction were quantified by placing the pendulum on a scale and measuring the change in apparent weight when a voltage was applied. Impulses ranging from ~1 μN-s to more than 1 mN-s were exerted on the pendulum by sending 10 V signals of varying pulse width to the voice coil. Displacements on the pendulum were measured for each impulse to generate a calibration curve (Fig. 7A-B).

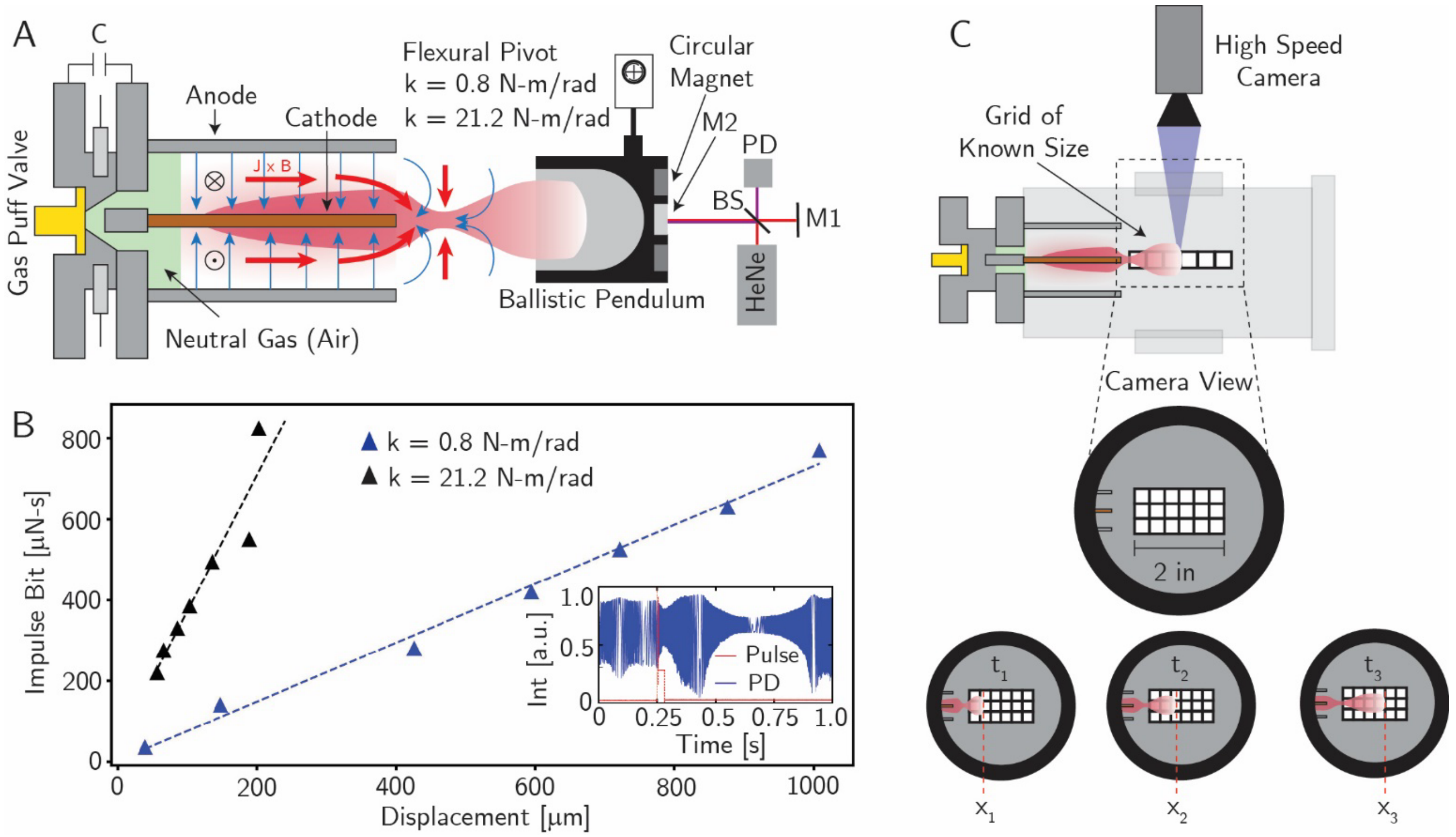

**Figure 7.** Experimental setup for the pendulum thrust stand that was used to measure impulse bits from the magneto-deflagration thruster. (A) Impulse bit was determined by measuring the displacement of a ballistic pendulum that was impacted by an accelerated plume. (B) The pendulum was calibrated by applying a known impulse and using a Michelson interferometer to measure displacements for two torsional stiffnesses. (C) A high-speed camera (10 million fps) used the time-of-flight method to infer exhaust velocity by tracking the leading edge of plumes as they expanded into vacuum.

### *3.4 Exhaust Velocity: Time-of-Flight*

A time-of-flight method was utilized to measure the exhaust velocity of plumes as they exited from the acceleration volume of the thruster. This method tracked the leading-edge interface of plumes over time as they expanded into vacuum to infer (1) peak velocities and (2) mass utilization (for a known $m_{bit}$) by comparing with $I_{sp}$ measurements from the pendulum thrust stand. Measurements were made by visualizing the broadband self-emission (integrated across the visible spectrum) from the plasma with a Shimadzu HPV-X1 high speed camera. The recording speed of the camera was set to ten million frames per second (fps) and was triggered by the capacitor discharge. After triggering, the camera took an image every μs for 256 μs to capture a full firing event. Images were timestamped during the recording and coupled with a spatial calibration grid to identify the axial position of the plume front across different frames. A spatial grid measuring 2.5 cm x 5 cm was placed along the centerline of the thruster to calibrate the camera spatially (Fig. 7C). Each square pixel was calculated to have side lengths of ~0.3 mm. Movement of the plasma plume across pixels from high-speed imaging was then correlated to expansion of the plume in real space using the spatial calibration. Exhaust velocity was measured by fitting a line to the leading-edge position of the plume across time.

## 4. Results & Discussion

### *4.1 MHD Mode Characteristics and Regime Boundaries*

Experiments were performed to identify regime boundaries between MHD operating modes in a pulsed electromagnetic accelerator, including conditions that lead to transitions between the magneto-detonation and magneto-deflagration modes (Fig. 8A). Initial conditions within the coaxial acceleration volume were varied to investigate how they influenced the evolution of ionization waves, the structure of plumes, the MHD mode of operation, and the properties of the thruster (e.g., exhaust velocity, T/P, etc.). Each experiment tracked the plume structure as it exited the acceleration volume under varying peak currents (I ~ 6-14 kA) and downstream propellant mass (i.e., air). The distribution of air in the accelerator at the moment an ionization wave formed was adjusted ($m_{dwn}$ ~20–130 $\mu$g) by varying the plenum pressure of the gas puff valve. To generate discharges with $m_{dwn}$ < 80 μg, a DC pre-ionization circuit was used to introduce seed electrons into the acceleration volume. High-speed broadband imaging of plasma self-emission was used to identify key features of the transient acceleration process as plumes exited the thruster, including the plume structure, collimation, and the presence of macroscopic instabilities near the exit plane of the thruster. Mode 1 (i.e., magneto-detonation) was characterized by a current sheet and shock wave propagating into the vacuum that was followed by an expansion wave, while Mode 2 (i.e., magneto-deflagration) produced a smoothly expanding, collimated plasma jet that was both quasi-steady and accompanied with a radial compression region, analogous to a magnetic nozzle, at the exit plane (Fig. 8B). Additionally, unstable transitions between these modes were observed, marked by the presence of macroscopic instabilities that led to periodic disassembly of the plasma jet.

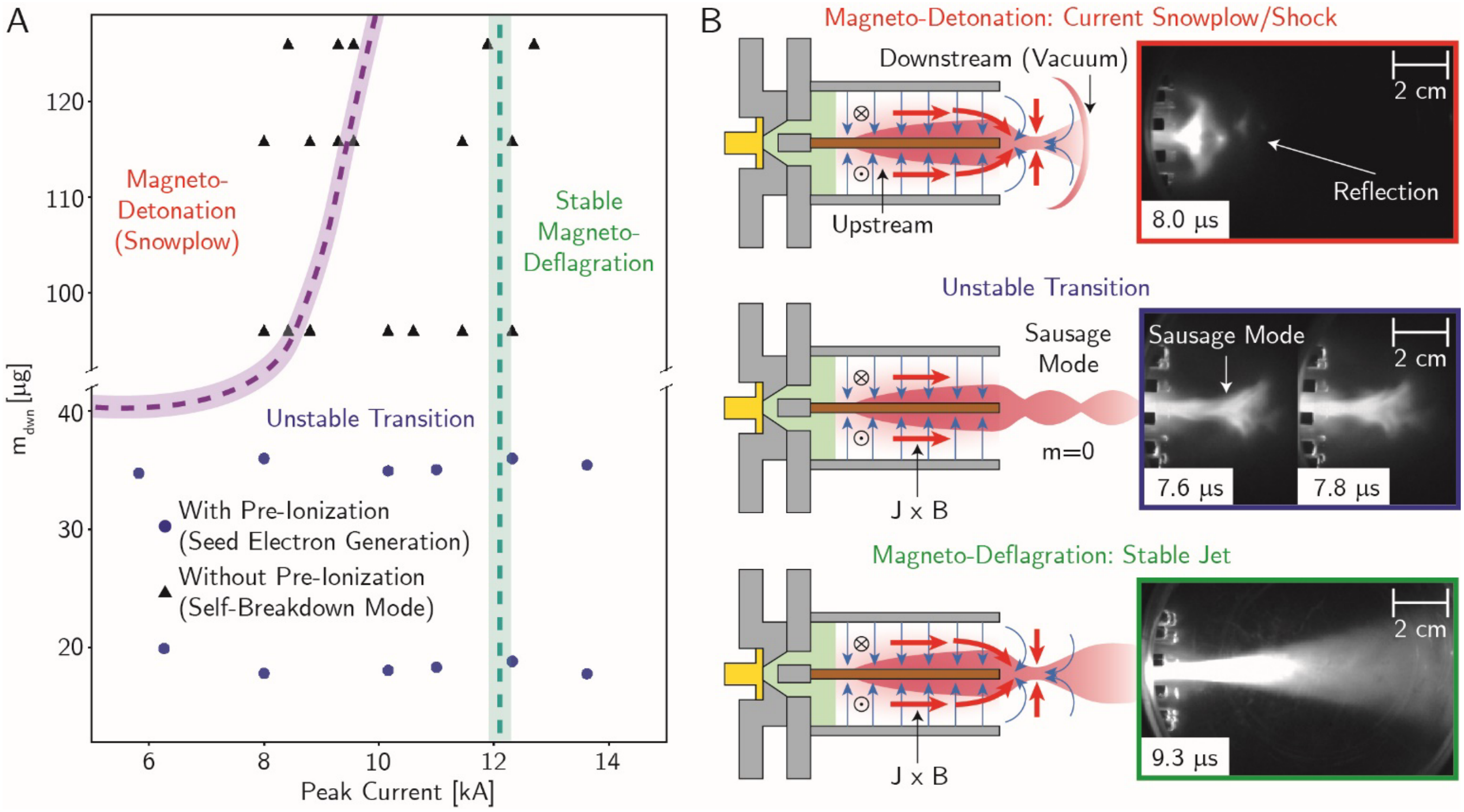

**Figure 8.** (A) Gas-fed electromagnetic accelerators operate in three distinct regimes during transient conditions: a magneto-detonation mode, a magneto-deflagration mode, and an unstable transition between them. (B) Regime boundaries highlight how the conditions within a thruster influence the mode of operation. These conditions were determined by tracking the structure and properties (i.e., exhaust velocity, impulse bit) of the thruster as operating conditions were varied. Stable magneto-deflagration waves occur above a threshold current (12 kA) for self-breakdown conditions regardless of $m_{ups}$. Increasing downstream propellant causes a mode to transition to a magneto-detonation (or snowplow) wave. This process occurs as gas temperatures and electrical conductivities increase locally within a shock. Operating the thruster at lower peak currents introduces magnetohydrodynamic instabilities that can disassemble plasma jets and induce transitions between modes.

A regime map (Fig. 8) summarizes the boundaries between MHD operating modes in PPTs, including the sensitivity to both operating current and propellant loading. Each point in the map represents a separate combination of initial conditions that the thruster was operated at (26 in total). Broadband imaging of the plume structure at 10 MHz resolved a number of distinct transitions including boundaries where plumes exhibited shock fronts as they convected from the acceleration volume, a smooth and collimated transition into vacuum, and macroscopic MHD instabilities. Experiments found that a combination of low peak current (I < 9 kA) and high downstream mass ($m_{dwn}$ > 40 μg) resulted in magneto-detonation waves. Broadband self-emission images revealed that the leading edge of the magneto-detonation wave formed a distinct shock front that curved backwards toward the electrodes, sketching current streamlines at the exit plane (Fig. 8B). Conversely, for higher peak current (I > 12 kA), a stable magneto-deflagration was observed across all values of $m_{dwn}$ that were tested. Between these two regimes, an intermediate "unstable transition" regime was identified at 9 kA < I < 12 kA. This regime was characterized by the presence of magnetohydrodynamic instabilities (e.g., sausage and kink modes) at the outer radius of the plasma column (Fig. 8B). Plasma in this regime was partially collimated and exhibited periodic, unstable ejection rather than the steady acceleration seen in Mode 2. The transition regime could also be induced at lower currents by using pre-ionization, which reduced the amount of neutral propellant available before breakdown.

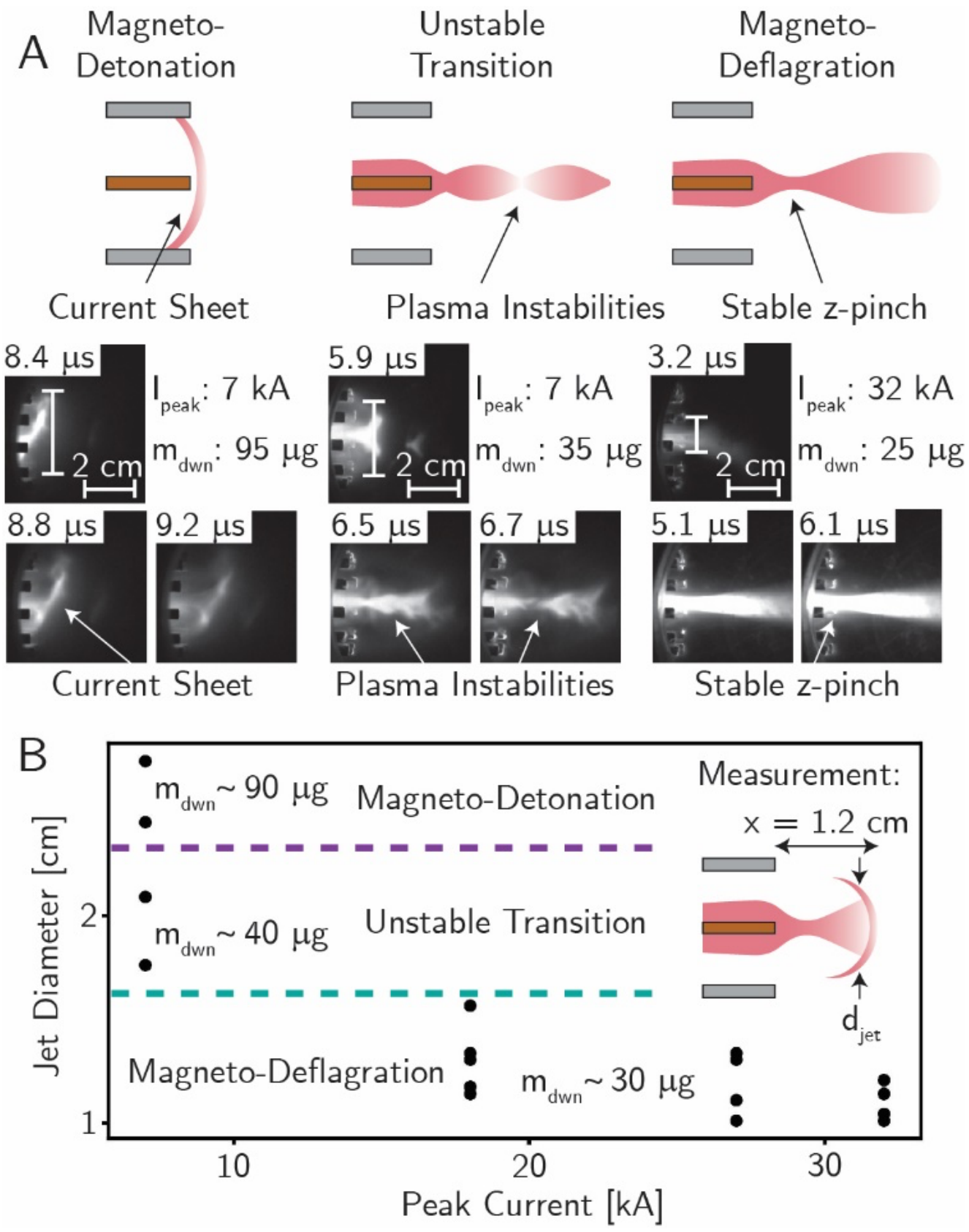

**Figure 9.** (A) The MHD operating mode was determined in part by tracking the structure of plumes as they exited the accelerator using high-speed (10 million fps) broadband imaging. Thruster operation was categorized into three modes based on key characteristics, including the plume radius (and the presence of a shock front), presence of macroscopic instabilities, and degree of collimation. (B) Plumes were observed to have smaller radii, increasing collimation, and decreasing evidence of instabilities as peak current was increased. Plumes were observed to feature shock fronts as more propellant mass was injected. Plasma radius measurements were taken 1.2 cm downstream from the cathode tip.

Different operating modes also exhibited clear differences in the plasma travel time and plume diameter downstream of the accelerator (Fig. 9A). For magneto-detonations it took > 8 μs for the plasma to exit the acceleration volume after breakdown occurred; in comparison, magneto-deflagrations took ~ 3 μs. Jet diameter was found to be more sensitive to downstream mass at lower discharge energies. At 6 kA, magneto-detonation fronts varied widely in size (~1.8–2.3 cm) as $m_{dwn}$ increased (Fig. 9B). In contrast, at 32 kA, magneto-deflagration jets remained highly stable, maintaining a diameter of 1.1 ± 0.1 cm regardless of downstream mass. As more current was used, the thruster began to operate exclusively in the magneto-deflagration regime, with more current driving additional radial compression at the exit plane and a higher degree of collimation (from ~1.3 cm at 18 kA down to ~1.05 cm at 32 kA, (Fig. 9B)).

Experiments found that different initial conditions within a pulsed electromagnetic thruster changed its MHD operating mode and the structure of the acceleration process. Visualizations of the plume found that two stable modes were observed, consistent with the RH theory, along with an unstable transition that connects them. The observed trends in jet diameter and plume travel time reinforce the distinct characteristics of each mode. The identification of an intermediate unstable transition regime, where plasma exhibits MHD instabilities and partial collimation, highlights the sensitivity of wave formation to accelerator feed conditions, including how neutral propellant is distributed initially in the volume. These dependencies suggest a mechanism to control the operating mode of PPTs on-demand, enabling operational flexibility based on mission requirements.

### *4.2 Mass Utilization within the Magneto-Deflagration Mode*

Exhaust velocity and impulse measurements were conducted for various MHD operating modes in the coaxial accelerator. A thrust stand and time-of-flight imaging were used to evaluate the performance of each discharge process under identical input energy and peak current conditions. These measurements aimed to identify differences in the acceleration mechanisms of each MHD mode, including propellant utilization, T/P, exhaust velocity, and $I_{sp}$, while maintaining the same input energy. Magneto-detonation waves were generated by disabling the pre-ionization circuit, which increased the downstream propellant mass without altering the charging voltage or propellant injection process.

Time-of-flight measurements of weakly unstable magneto-deflagration modes (i.e., in the unstable transition regime) indicated that their exhaust velocities were twice as high as those of magneto-detonation modes with identical input energy and peak current. These velocities were determined by tracking the leading edge of the plume as it expanded into a vacuum over six frames at 10 MHz (Fig. 10A). With pre-ionization enabled, exhaust velocities of approximately 55 km/s were recorded at 7 kA with a downstream mass of ~60 μg (Fig. 10B-C). When pre-ionization was disabled—while keeping the charging voltage and valve conditions unchanged—the downstream mass increased to ~120 μg, resulting in a proportional decrease in exhaust velocity to ~20 km/s. This decrease in velocity was accompanied by an apparent change in the structure of the plume, from an unstable magneto-deflagration to a magneto-detonation mode.

Experiments were also conducted to measure the propellant utilization efficiency $\eta_{util}$ of the coaxial accelerator as it transitioned into the magneto-deflagration regime. These measurements were taken while the accelerator operated at a fixed charging energy and peak current (~8 kA), with variations in the propellant distribution. Valve opening times ranged from 800 to 2000 μs, and injection pressures were adjusted between 0 and 40 psig, resulting in mass bits from 10 to 400 μg and an estimated downstream mass of 10 to 70 μg across different discharges. Pre-ionization was applied to ensure operation in the unstable magneto-deflagration regime at this discharge current. The propellant utilization efficiency was determined by calculating the specific impulse as the ratio $I_{sp} = I_{bit}/(g_0 m_{dwn})$, where $I_{bit}$ represents the measured impulse bit and $m_{dwn}$ is the downstream injected propellant mass bit during a single discharge event. This was compared with the exhaust velocity inferred from time-of-flight analysis for a given charging energy. Differences in the measured impulse bit from the time-of-flight exhaust velocity and known mass bit indicate the fraction of propellant that remains unutilized in the acceleration volume, with $\eta_{util}$ given by $\eta_{util} = I_{sp}g_0/v_{ex,TOF}$.

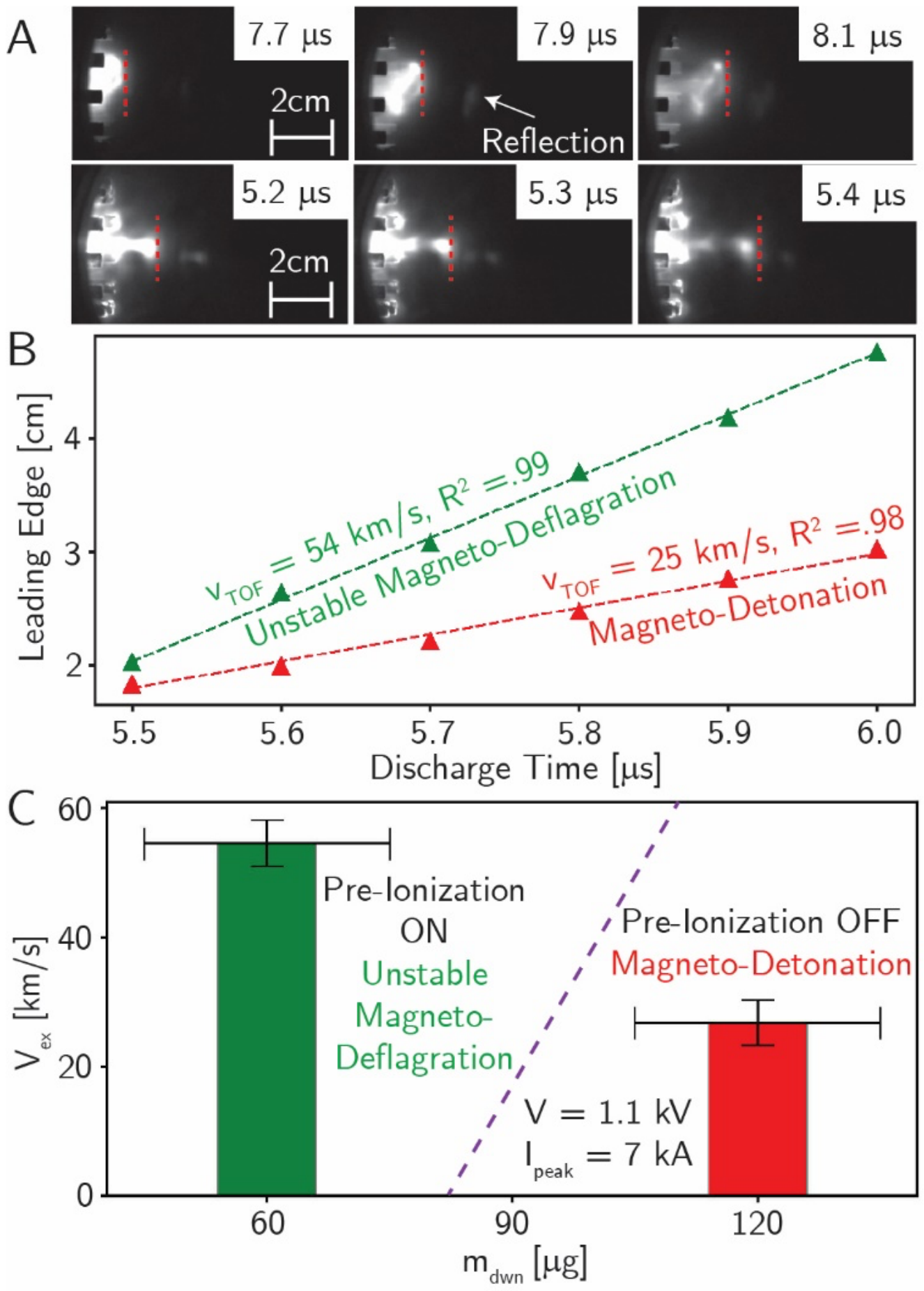

**Figure 10.** (A) The leading edge of the plasma was tracked over time for a fixed charging energy (28 J) and different propellant loadings ($m_{ups}$ = 60 µg and $m_{ups}$ = 120 µg) within the electromagnetic accelerator. (B) The exhaust velocity of each condition was determined using time-of-flight tracking of the plume's leading edge. Timestamps correspond to the discharge time. (C) The exhaust velocity of the magneto-deflagration mode was measured to be two times higher than that of magneto-detonation mode.

Measurements of propellant utilization in the magneto-deflagration regime, including in conditions where instabilities were present, revealed several trends in the performance of the accelerator. When operated at a fixed peak current while varying the propellant loading (i.e., downstream mass), the time-of-flight exhaust velocity remained constant at ~55 km/s (Fig. 11A). This suggests that the peak velocity and acceleration process remained unchanged as additional downstream propellant was introduced—unlike in the magneto-detonation regime, where structural changes in the flow were accompanied by a halving of the exhaust velocity (Fig. 10). However, thrust stand measurements of $I_{bit}$ varied from 334 µN-s to 140 µN-s as the downstream propellant mass ranged from 70 to 15 µg in the magneto-deflagration regime. This indicates that the $I_{sp}$ increased with decreasing propellant injection, ranging from 100 to 2000 s (Fig. 11B). Similarly, propellant utilization efficiency increased from 5% to 35% as the amount of injected propellant decreased (Fig. 11B). This indicates that the magneto-deflagration mode becomes more efficient as less propellant is downstream of the ionization wave, as a greater fraction of the available mass remains entrained in the plume. This trend also applies to the total injected propellant mass, including the mass still present in the volume between the valve and the acceleration region. As less total mass is injected, a greater fraction is within the acceleration volume when a discharge forms, allowing it to be accelerated into the plume to generate thrust. Any mass that does not make it into the acceleration volume when a discharge begins remains unutilized. These trends continued until the downstream mass bit of air approached ~ 15 µg. At this point, for a charging energy of 40 J, the system transitioned into underfed conditions, where ablative effects began to play a role. This transition is characterized by the condition where the number of electrons driven in the discharge waveform ($n_e$) exceeds the number of injected gas molecules in the propellant stream ($n_g$) such that $n_e > n_g$ (Fig. 11C). Beyond this point, the mass contribution includes not only the injected propellant but also material ablated from the electrodes. Further experiments to quantify this transition are described in Section 4.3.

The results provide new insights into how different MHD modes affect the performance of PPT. For a fixed energy condition, the magneto-deflagration regime exhibits higher exhaust velocities and the potential for higher propellant utilization as less propellant is injected. Measurements also highlight the role of initial conditions —such as the distribution of propellant and timescale of energy addition—in determining performance of electromagnetic thrusters. These findings are particularly significant for PPTs, which have suffered from poor mass utilization traditionally. By demonstrating that performance can be enhanced through precise control of discharge pulse shapes, timescales, and propellant gas dynamics, these results establish the viability of PPTs for applications such as ABEP, where efficient propellant utilization is essential.

### *4.3 Thrust Scaling in the Magneto-Deflagration Mode*

Measurements were performed to evaluate the scaling of PPT performance that operate in the magneto-deflagration regime (Fig. 12). Specifically, experiments examined how thrust in this regime scaled with increasing input current and charging energy, with the aim of understanding how the acceleration mechanism changed as more current flowed within the acceleration volume. A thrust stand was used to assess these changes by measuring the $I_{bit}$ of a coaxial electromagnetic accelerator for two different propellant loadings (25 µg and 50 μg) as charging energy varied from approximately 20–230 J, corresponding to peak currents of 5-22 kA. For all currents considered, both propellant loadings remained within the magneto-deflagration regime boundary shown in Fig. 8A. These measurements were taken on a single-shot basis and were also analyzed using T/P by combining measurements of $I_{bit}$ and input energy for individual discharge events.

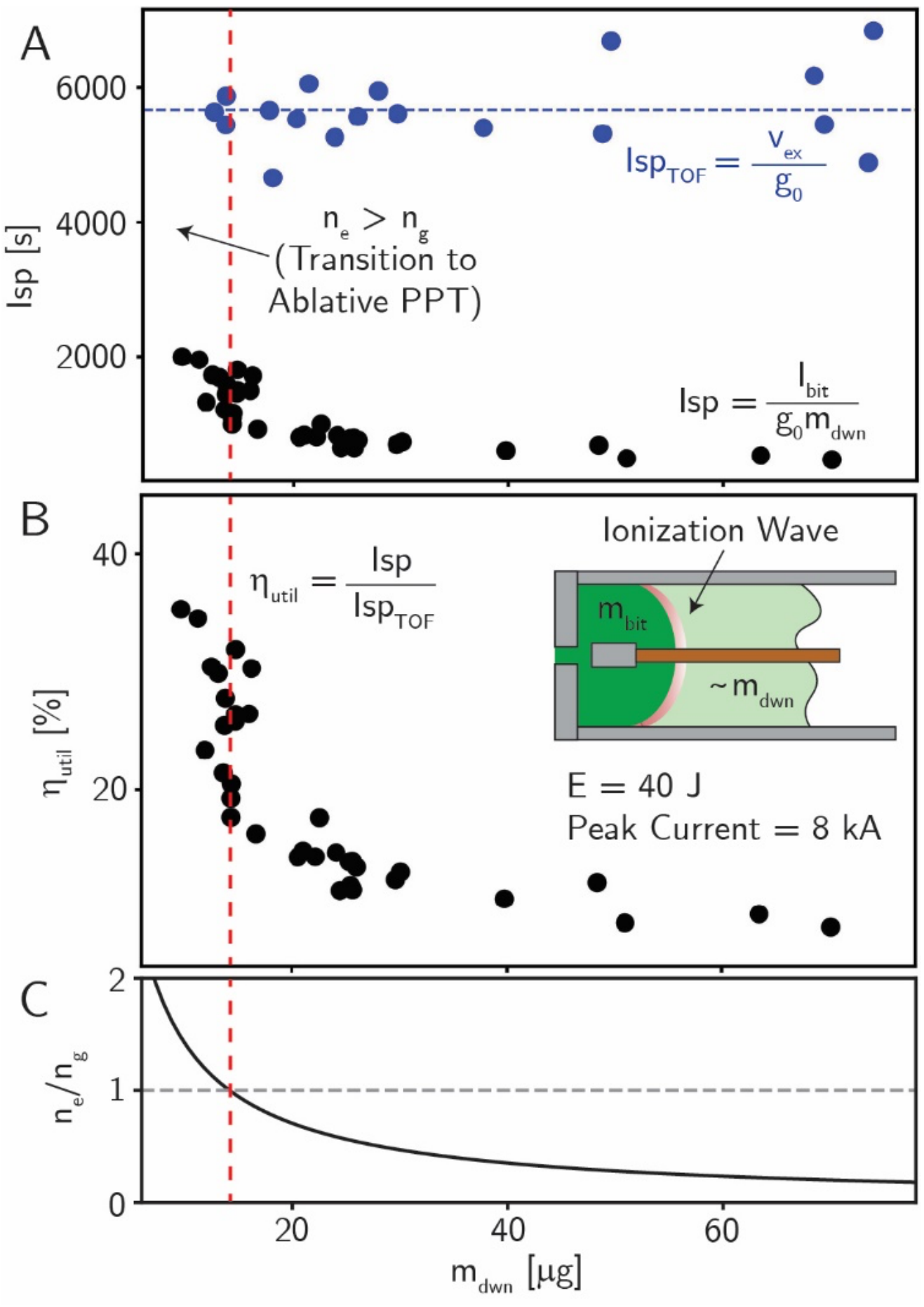

**Figure 11.** (A) The magneto-deflagration and unstable transition modes scale with the amount of propellant that is injected into an accelerator for a fixed amount of energy. (A) Measurements of $I_{sp}$ for the thruster operating in a magneto-deflagration/unstable transition mode for a fixed charging energy of 40 J (8 kA peak current). The maximum $I_{sp}$ of the plume, inferred from the exhaust velocity, remains unchanged for all values of $m_{ups}$, indicating the thruster remains in the same mode. However, the inferred $I_{sp}$ using the measured $I_{bit}$ increases for decreasing $m_{ups}$, suggesting that a greater fraction of the injected propellant is entrained in the ionization wave. (B) This results in an increase in $\eta_{util}$ as less propellant is introduced to the accelerator. (C) The sharp growth in $\eta_{util}$ when $m_{ups}$ ~15 μg could be influenced by a transition to ablative operation, where the number of electrons driven by the discharge ($n_e$) exceeds the number of gas molecules introduced into the thruster ($n_g$).

The $I_{bit}$ of the magneto-deflagration mode was found to increase from 100 to more than 1700 μN-s per discharge (Fig. 12A) as charging energy and $\int I^2\, dt$ increased (Fig. 12B). This trend remained consistent across different propellant injections, including total injected masses of approximately $m_{bit,1} \approx 25$ μg and $m_{bit,2} \approx 50$ μg. This scaling behavior was expected, as $\int I^2\, dt$ is proportional to the J×B force driving an induced-field electromagnetic thruster, and it held for $\int I^2\, dt \lesssim 1800$ A²s (for 50 μg) and $\int I^2\, dt \lesssim 1600$ A²s (for 25 μg) (Fig. 12B). However, above these thresholds, a transition from linear to quadratic scaling of $I_{bit}$ was measured. In the linear regime, $I_{bit}$ scaled identically for both propellant loadings. However, in the quadratic regime, discharges with 25 μg of propellant exhibited an increase to ~1750 μN-s, nearly 20% more than the highest impulse bit with 50 μg injections for $\int I^2\, dt = 4000$ A²s.

Thrust-to-power measurements were quantified by combining $I_{bit}$ and input energy measurements for single discharges. T/P serves as an indicator of acceleration efficiency, where a proportional increase in $I_{bit}$ and discharge energy results in a constant T/P value. Measurements in the magneto-deflagration regime revealed the presence of both a constant and linear scaling regimes, depending on the total amount of injected propellant. In the constant regime, T/P remained at approximately 4.5 mN/kW for both propellant loadings, suggesting that a larger fraction of the injected propellant did not reach the acceleration volume at higher injection conditions. This is expected, as the Paschen limit remains unchanged for a given charging energy, meaning any mass upstream of the ionization wave remains unutilized.

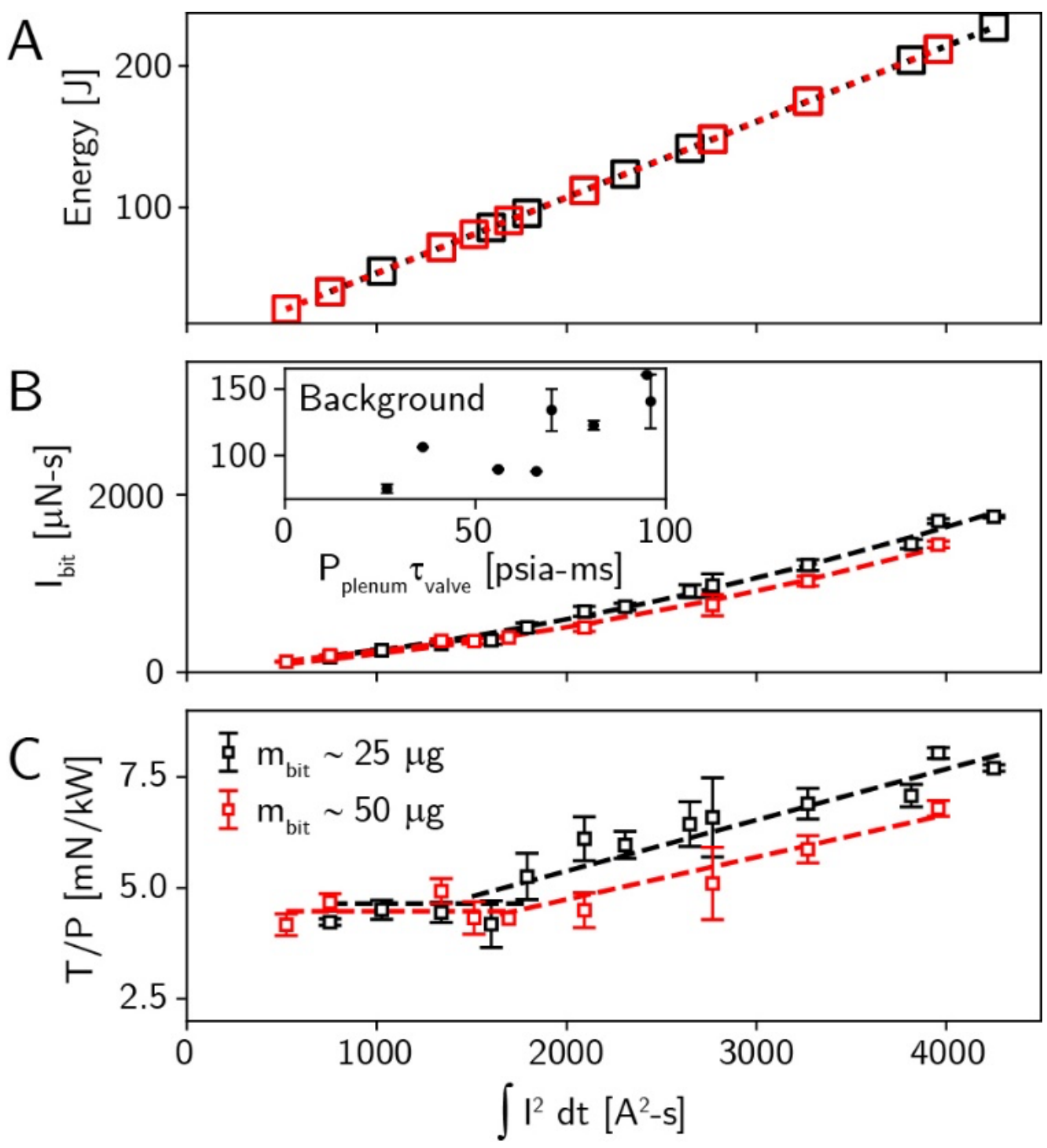

**Figure 12.** (A-B) The $I_{bit}$ of the magneto-deflagration mode scales with charging energy in two different regimes, (1) linearly for $\int I^2\, dt \lesssim 1800$ A²s and (2) quadratically for $\int I^2\, dt > 1800$ A²s (for $m_{bit}$ ~50 μg). These regimes quantify the scaling of the JxB force, which for a self-field thruster, increases as $\int I^2\, dt$. (C) Thrust to power ratios, equivalent to $I_{bit}$/charging energy for a single shot, transitions from a constant value when $\int I^2 \lesssim 1800$ A²s to linear growth in T/P when $\int I^2 > 1800$ A²s. The transitions shift to higher values of $\int I^2\, dt$ when more propellant is injected.

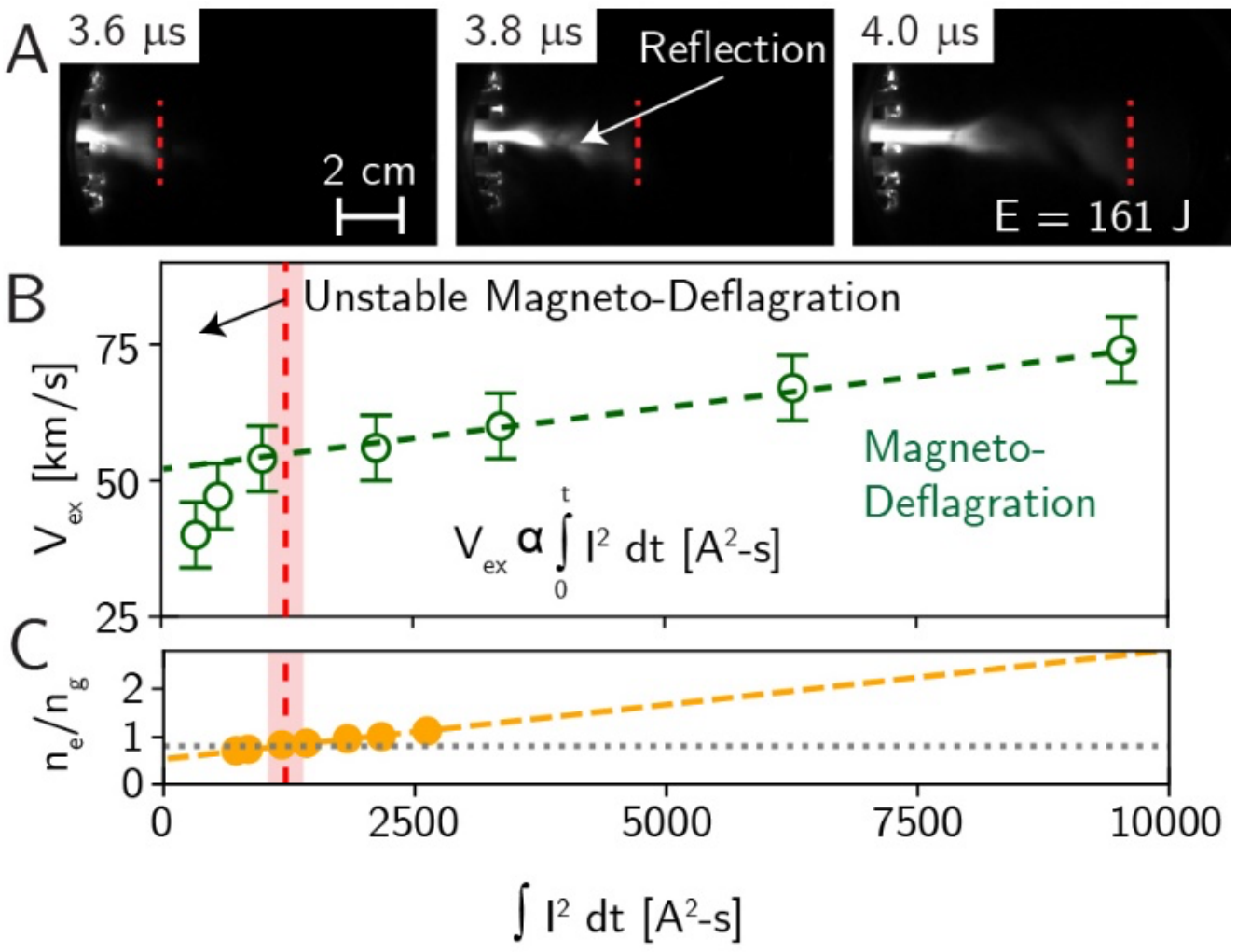

**Figure 13.** (A) TOF image sequence of the magneto-deflagration jet with an input current of 18 kA (161 J). (B) Exhaust velocity increases sharply for $\int I^2 < 1600$ A$^2$s where the magneto-deflagration is unstable. (C) The stable magneto-deflagration mode features an exhaust velocity that increases linearly with increasing $\int I^2$ dt. This includes the regime $\int I^2 > 1600$ A$^2$s where the $I_{bit}$ scales quadratically, supporting the hypothesis that increasing mass from ablation contributes to the nonlinear scaling of thrust. (B) The transition to ablative effects can be estimated when $n_e/n_g > 1$.

Above thresholds in current, the T/P of the thruster began to increase linearly for both propellant injections. From 1800 A$^2$s $\lesssim \int I^2 \lesssim$ 4000 A$^2$s, $m_{bit,1}$ increased from ~4.5 mN/kW to ~8 mN/kW and $m_{bit,2}$ increased from ~4.5 mN/kW to ~6.7 mN/kW.

An increasing T/P ratio suggests that the accelerator's performance can be tuned and that the underlying acceleration mechanisms have become more efficient. To identify the mechanism responsible for this improvement in the magneto-deflagration mode—and to understand why a transition occurs at specific operating currents—additional experiments were conducted using time-of-flight broadband imaging. These measurements were used to quantify how $v_{ex}$ scales under different operating conditions. Time-of-flight diagnostics were used to examine how exhaust velocity evolved as the jet transitioned into the magneto-deflagration mode, providing insight into the observed nonlinear trends in $I_{bit}$ and T/P. Measurements began at an input current of 7 kA, within the unstable transition regime. Current was then increased incrementally to allow a transition into the magneto-deflagration regime. A representative time-of-flight case at 18 kA is shown in Fig. 13A, where elapsed time stamps indicate that at this current, the plume exits the thruster approximately 7.5 μs after discharge initiation. The results show that for $\int I^2$ dt $\lesssim$ 1600 A²s, $v_{ex}$ increases sharply from ~35 km/s to 55 km/s as the thruster enters the stable magneto-deflagration regime. Beyond $\int I^2$ dt = 1600 A²s, $v_{ex}$ scales linearly with $\int I^2$ dt. Since impulse bit is given by $I_{bit} = \eta_{util} m_{bit} v_{ex}$, a linear increase in $v_{ex}$ with $\int I^2$ dt would imply that T/P = $I_{bit}$/Energy should remain constant for the same propellant loading. However, measurements show that T/P continues to increase linearly beyond $\int I^2$ dt = 1600 A²s (Fig. 12B), indicating that the transition is driven by changes in mass utilization efficiency or the availability of propellant mass.

To investigate the cause of increasing T/P and propellant mass at higher charging conditions, the number of injected neutral gas particles $n_g$ with the number of electrons $n_e$ carrying the discharge current were calculated for each firing condition in the magneto-deflagration mode. The number of electrons and the total number of injected gas molecules were calculated by integrating the capacitor current traces and measuring the injected $m_{bit}$, respectfully, for each data point. These experiments featured two separate amounts of injected propellant across many different firing conditions at elevated charging energies, so transitions between operation were dictated by the amount of current that was driven through the propellant. For optimal PPT performance, neutral gas density should be sufficient to sustain current flow, preventing plasma from retreating toward the breech during discharge or extracting mass from electrodes *(58)*. When $n_e/n_g > 1$ the accelerator enters a "starved" feed condition, forcing the plasma to propagate toward the breech in search of charge carriers. In extreme cases, the plasma ablates material from the thruster itself. Scaling experiments caused the accelerator to enter the starved regime at $\int I^2 \sim 1600$ A$^2$s for $m_{bit,1} \approx 25$ μg (Fig. 13C). Beyond this threshold, the number of electrons in the discharge exceeded the number of available neutral molecules. With this context, the scaling trends in Fig. 12B-C can be reinterpreted. In the overfed regime, $I_{bit}$ increases linearly. However, upon entering the starved regime, mass ablation from the device likely becomes significant, causing the entrained $m_{bit}$ to scale with increasing current. This transition could explain the quadratic scaling of $I_{bit}$ and accounts for the earlier onset of quadratic behavior observed for lower injected propellant masses ($m_{bit,1} \approx 25$ μg).

## 5. Conclusions

This study explored the formation and tunability of MHD modes in pulsed plasma thrusters, addressing a key knowledge gap in how these acceleration mechanisms emerge and transition between regimes. By varying propellant gas dynamics, pulse energy, and discharge timing systematically, we demonstrated the ability to control ionization wave behavior and induce transitions between acceleration modes. Our results show that the initial conditions of ionization waves within the acceleration volume dictate the structure and characteristics of the acceleration process. Time-of-flight and thrust stand diagnostics revealed that in the magneto-deflagration mode, exhaust velocity scales linearly with input energy, while $I_{bit}$ and T/P exhibit nonlinear transitions at a critical current threshold. Furthermore, the study quantified improvements in propellant utilization efficiency, showing that reducing injected mass enhances mass entrainment within the deflagration wave. This effect led to a 20× increase in inferred $I_{sp}$, demonstrating a pathway to optimize performance by controlling downstream mass injection.

These findings provide new strategies for optimizing electromagnetic acceleration in PPTs, particularly for applications requiring high $I_{sp}$ and efficient mass utilization. The demonstrated improvements in propellant efficiency address a longstanding limitation of PPTs, enhancing their viability for ABEP and other low-mass-flow missions. More broadly, the ability to tune acceleration modes through controllable operating parameters or pulse shaping introduces new design possibilities for plasma-based propulsion systems, increasing the operational flexibility of PPTs and other electromagnetic thrusters. The insights gained into propellant gas dynamics and efficiency scaling within distinct acceleration structures lay the groundwork for further advancements in electromagnetic propulsion architectures, with potential implications for high-efficiency space propulsion, material processing, and controlled plasma confinement applications.

**Acknowledgements**

This research is supported by the Defense Advanced Research Projects Agency TALOS program award UTAUS-FA00003150 with Dr. Susan Swithenbank as Program Manager.